\documentclass[aps,prd,reprint,showpacs,showkeys,amsmath,amssymb,amsfonts]{revtex4-1}
\pdfoutput=1
\usepackage{blindtext}
\usepackage{graphicx}
\usepackage{fullpage}
\usepackage{dcolumn}
\usepackage{hyperref}
\usepackage{tabularx}
\usepackage{natbib}
\usepackage{mathtools}
\usepackage[x11names]{xcolor}
\usepackage{hyperref}
\hypersetup{
	allbordercolors={blue},
	colorlinks=false, 
	citecolor={blue},	 
	citebordercolor={blue},
	linktoc=all,	 
	linkcolor={blue},	 
	urlbordercolor={Gold1}
}

\begin{document}
\title{Scalar-Tensor Gravitational Strain Field Equations and the
Longitudinal Wave Form}
\author{Ronald S. Gamble, Jr}
\email{rsgamble@ncat.edu}
\author{K. M. Flurchick}
\affiliation{\\North Carolina Agricultural \& Technical State
University\\Greensboro, NC, 27401}

\begin{abstract}
From modern observations of gravitational interactions, it can be inferred that
there is much left to discover about the fundamental gravitational field. Since
the advent of the General Theory of Relativity over a century ago, we have come
to make exotic assumptions pertaining to the inner workings of an associated
field theory. One of which is an elastic nature to spacetime and the behavior of
gravity for strong and weak fields. In this work we investigate a more \textit{physical}
nature, expanding upon general relativity led by observations of strong sources.
We introduce a candidate Lorentz-invariant field theory that employs an
\textit{elastic} and \textit{pseudoscalar} nature to the field interpretation
and it's properties.
A unique generation of the Euler-Lagrange equations of motion is presented; resulting in
a longitudinal wave equation for the \textit{Dilation} gravitational field. This
provides a modern advancement of a relativistic gravitational field theory,
supported by observation.

\end{abstract}


\maketitle

\section{Introduction}

At the advent of the 20th century research conducted on fundamental physics
generated a vast growth of the understanding of principal physical phenomena
like that of Electromagnetism, Gravitation, Nuclear Physics and the genesis of
Quantum Mechanics. Today, investigations into the mechanisms of the fundamental
forces has taken a more robust approach contrasting older methods of research.
Physics beyond Feynman's introduction to modern unified field theory, in the
form of QED, changed our treatment of investigating candidate field theories of
fundamental forces. Historically, gravitational field theory has had few
fundamentally profound breakthroughs when compared to the progress of other
fundamental forces (strong and weak nuclear, and electromagnetism).  Leaving a
gaping hole in our understanding of the force in regards to a \textit{dynamic}
or time-dependant nature; and subsequently a quantum description of the
gravitational field. Following with the communications of Einstein, Rosen and
Infeld \cite{hist2} from 1936-1938, Einstein himself was adamant that
gravitational waves were not physically manifest and were a mathematical
artifact of the linearized field equations per his publications in
1916-1918\cite{hist1,hist0}.
In which [he], Infeld and Hoffman arrived at the \textit{post-Newtonian
approximation} to the weak field equations \cite{hist2} generating a perturbed
\textit{linear} set of plane wave equations in cylindrical coordinates; but not
without apparent coordinate singularities propagated by these cylindrical plane
waves. For many years the theoretical development of gravitational waves has
eluded many scientists. Progressing from the beginning conversations exchanged
between Einstein, Rosen and Infeld debating the mathematical artifacts arising
from the implementation of a perturbation method\cite{hist2}, to the eventual
discovery made by the LIGO/LSC community of scientists\cite{ligo}. Experimental
and observational studies have only tested for and observed gravitational waves
to propagate transversely through space.
These spatially transverse waves constitute the formal oscillations allowed by
the well-known transverse-traceless gauge conditions\cite{ttgauge}. Within these
gauge conditions and under a linearization method, the reconstruction of the
Riemann curvature and subsequently the Ricci curvature fails under this regime.

The current relativistic theory of gravitation, \textit{The General Theory of
Relativity} (GR), was not constructed to admit sufficient time-dependent
solutions for the temporal variation of the gravitational field.
Nor does it admit a substantial theory of gravitational field oscillations
emanating as radiation in regards to localizing the energy-momentum carried. For
gravitation proposed as a fundamental spin-2 self-interacting field, an
appropriate nonlinear representation of the field equations must be
established. In order to recover the necessary curvature on a curved background,
one must associate a source term in the field equations with gravitational
mass-energy. Direct and indirect observations of strong sources (Black Holes,
Neutron Stars, Binary Mergers, etc.) and Dark Matter-Energy elude to us that
there is a missing facet to gravity not seen or theorized using GR under the
standard model. A wide variety of solutions, in GR and its extensions, have been
constructed to account for such dynamics introducing more complexities than
simplifications. Looking at the detailed detections of LIGO and similar
gravitational wave observatories, it is intuitive to infer that there is a
fundamental ``elastic'' nature to spacetime and the interactions involving
gravity. With this being said, in this paper we build upon the foundation of an
elastic interpretation of spacetime laid out in the previous work \cite{rgdiss}
and \cite{rg1}. We look into a formal description of volume deformations of
mass-energy distributions with respect to the scalar-tensor gravitational strain
field (\textit{Dilation}, $\mathcal{D}^{\alpha\beta}$).
When considering volume deformations of sourced mass distributions, this field (termed Dilation) constitutes deformations of spacetime
geometry when considering the principle directions or local volumes within a
local spacetime. 

Formulated from the observation of the aforementioned strong field sources, this
proposed pseudoscalar-like gravitational field extends the idea of static
curvature to a dynamic field that is scaled by the invariant curvature
$\Lambda^{\alpha\beta}R_{\alpha\beta}=R$ of the local space. Where,
$\Lambda^{\alpha\beta}$ is the strong-form metric of the local
spacetime\cite{rg1}. Mass densities and scalar curvatures are considered the sources of this field.
Under static conditions across $n=4$ spacetime dimensions this
field is defined by a finite volumetric gravitational strain tensor,
$\varepsilon^{\sigma\gamma}(\eta_{\mu},\tau)$, for a constant local mass
distribution. An expression for the dilation tensor is given in terms of the
finite volumetric strain and the trace of the Ricci curvature tensor.
\begin{equation} \label{dilation}
\mathcal{D}^{\mu\nu}(\eta_{\mu},\tau) =\frac{\beta
R_{vac}}{2m_{vac}}\Lambda^{\mu\nu}\left(\Lambda_{\sigma\gamma}\varepsilon^{\sigma\gamma}(\eta_{\mu},\tau)\right)
\end{equation}
In this expression the scalar curvature is provided by the Einstein field
equations of the strong-form background, in which the scalar holds
values for the background vacuum ($\Lambda_{\mu\nu}$) and the source
distribution.
Considering the ${}^{+}\Lambda$-vacuum as a zero state configuration, the
scalar curvature is always nonzero with a value of $R_{vac}\equiv R_{zero}$,
when $R_{zero} = \Lambda^{\alpha\sigma}R_{\alpha\sigma}\equiv 4({}^{+}\Lambda)$
for dimension (n = 4). For nonzero source distribution, $R_{vac} = R_{zero} +
R_{source}$ where $R_{source}$ can take on values from any solution of the trace
einstein field equations for nonzero mass. This expression for the dilation
field represents a way of dynamically describing the gravitational strain
generated by a nonzero mass-energy distribution using the trace Einstein field
equations; as first described in \cite{rg1}.

With this definition of the Dilation field, the development of a formal
Lagrangian density is given in section \ref{lagrangian} in which terms of
$\mathcal{O}(\mathcal{D}^{2})$ and higher are deemed valuable to the theory.
This lagrangian density is thus a representation of the gravitational energy
density involved in an isotropic compression and expansion of the local
spacetime. In section \ref{fieldeqn}, an action is then uniquely varied with
respect to the field, $\mathcal{D}^{\mu\nu}$, and then the background
metric, $\Lambda^{\mu\nu}$. Resulting in the recovery of the tensorial einstein
field equations and the corresponding field equation for dilation. Lastly
generating a nonlinear longitudinal gravitational wave equation, describing the
propagation of this proposed massive pseudoscalar field.
In the succeeding sections, the curvature of the spacetime background are
incorporated into the dilation field equations producing solutions dependent on
the Einstein Field Equations when the curvature is not negligible; for the
regime $\eta _{\mu \nu }\to \Lambda_{\mu \nu }$ incorporating strong gravitational fields.

\section{Lagrangian Density Of The Dilation Field}\label{lagrangian}

\subsection{The Massive Phonon Basis}
Looking into a fundamental basis of which we can formulate gravitational strain
field equations from, we arrive at the generalized form of the Klein-Gordon equation
representing massive free-scalar fields. Where the classical Klein-Gordon
equation can be expressed in terms a massive field, $\phi(x_{\mu},\tau)$, as
\begin{equation}\label{kg}
\left(\square + m^{2}\right)\phi(x_{\mu},\tau)=0
\end{equation}
Traditionally, the linear Klein-Gordon equation is utilized as a template
relativistic wave equation. In which it is commonly used for the introduction of
novel techniques and/or concepts pertaining to massive scalar or pseudoscalar
free-fields. Here we explore the implications of beginning with a classically
non-interacting massive scalar field. The description of this test field is
similar to that of a phonon with nonzero mass. We are thus directed to begin our
efforts with a generalized form of a lagrangian density describing the massive
phonon field (our test field), $\phi(x_{\mu},\tau)$. The Lagrangian density in a
lattice space, ($\mathcal{L}_{\text{phonon}}$), provides us with a 
statement of the associated energy density: $\mathcal{L}_{\text{phonon}}=\sum
_{k=1}^n \frac{m}{2}\left(\partial_\tau\phi \right){}^2-\frac{k_s}{2}\left(\phi
_{k+1}-\phi _k\right){}^2$
\begin{equation}
\mathcal{L}_{\text{phonon}}=\sum _{k=1}^n \frac{m}{2}\left(\partial_\tau\phi
\right){}^2-\frac{k_s}{2}\left(\phi _{k+1}-\phi _k\right){}^2
\end{equation}
Here $k$ is an index associated with the number of neighboring parcels of
spacetime, in which the field at that point has mass $m$. Under this discrete
approximation we make the assumption that the field is partitioned with respect
to an inter-parcel spacing $A$. Indicating that there is a spatial
discreteness about the test field at the $k^{th}$ point, $\phi_{k}(x)$, and its
nearest neighbor $(\phi_{k+1}(x) - \phi_{k}(x))$. This makes for a valid
approximation for a ``spacetime phonon'' in a discrete limit; behaving very much
like the classical treatment of traditional phonons in a 1-dimensional crystal
lattice. Continuing this analogy, in the continuum limit, where the sum is
replaced by an integral over space \(\sum _{k=1}^n\)$\to $ \(\frac{1}{A}\)$\int
dx$ and using a Taylor expansion on the field variable ($\phi $),
$\phi_k\to \sqrt{A}\phi \left(x_i\right)|_{x = k*A}$
and $\left(\phi _{k+1}-\phi _k\right) \to \sqrt{A^3}\nabla_i\phi
\left(x_i\right)|_{x = k*A}$; the corresponding lagrangian in
this limit is
\begin{equation}
\mathcal{L}\left(\phi,\partial_i\phi\right)=\frac{m}{2}\left(\partial _t\phi \right){}^2-\frac{k_s}{2}A^2\left(\nabla _i\phi \right){}^2
\end{equation}
as A $\to $ 0. We can see that the new behavior of this test field, again,
resembles that of a ``sapcetime phonon'' in a continuum limit. Conceptualizing a
vibratory nature to spacetime parcels. In comparison this gives a continuous
description of neighboring spacetime parcels as opposed to a discrete method
pertaining to discrete infinitesimal sectors of spacetime. The linear behavior
of this test field is analogous to that of the non-interacting longitudinal
gravitational strain field (Dilation). The full interacting field description is
explained in further detail following this section.

Fundamentally, a linear longitudinal oscillation is comprised of
displacements ($\Delta S_{\mu}$) parallel to the direction of propagation
resulting in the relaxation (expansion) and rarefaction (compression) of a local
volume. Notably, one might be inclined to suggest from this analogy that the
Klein-Gordon equation alone provides a simple expression that may be used in the
development of a resulting field equation. Suggesting that modifications could
be made to the equation to condition or fashion it for use within the framework of
general relativity directly. Unfortunately, the Klein-Gordon equation alone does
not hold the necessary structure to cater to a scalar-tensor gravitational
field. Lacking the necessary nonlinearity predicted by a generalization of
interacting scalar-tensor fields. Thus, describing the tensor field of interest
as a free-scalar is insufficient for the description of time-dependent
variations in volume deformations of mass-energy distributions. We must move to
consider an interacting field theory that includes nonlinear terms involving the
field variable $\mathcal{D}_{\mu\nu}$. With that being said, the Klein-Gordon
equation provides us with a valid linear approximation from which we can begin
the formulation of the formal Dilation, Gravitational Strain field equations.

\subsection{Lagrangian Density of the Dilation Field}

Let the tensor field $\mathcal{D}_{\mu \nu }$ be the object of interest. This
tensor is expressed in terms of the components of the volumetric gravitational
strain tensor $\varepsilon_{\mu\nu}$ in a 4-dimensional spacetime.
Corresponding to the principle directions of strain, this presents the Dilation
as a simple tensor with no off-diagonal elements. The trace $\Lambda ^{\mu \nu
}\mathcal{D}_{\mu \nu }\equiv 4\left| \mathcal{D}(\Lambda)\right|$ provides us
with a description of volume deformations of mass-energy distributions
\(\left(\rho _0\right)\) in a local spacetime. Looking back at equation
\ref{dilation}, this massive field effectively provides a means of transporting
the scalar curvature associated with a local spacetime, scaled by the
{``}resistance{''} or effective mass, $\mathit{m}_{\text{vac}}=\frac{\Lambda
c^2}{8\pi G}\int_v\sqrt{-|\Lambda| }d^{\mu }x_{\mu}$, of the background vacuum.
In regards to a field equation further explained in the succeeding section, we
consider the interaction of this field with the background of nonzero
effective mass. This generates the sought after nonlinearities incorporating the
Klein-Gordon form from the phonon propagation in
resistive media analogy in the preceding section. A massive longitudinal wave
equation can then be written from the nonlinear theory that would give the
parallel displacements of constituent finite spacetime parcels with zero spacing between them. Analogous to the model of a lattice phonon as the vibratory motion of a particle continuum, the Dilation field can be
described very much like the phononic model with appropriate modifications for a
\textit{classical} field\cite{scalar1,klein4}.
With this, spacetime fluidity can be described by a collection of local unit
volumes with mass $m_{vac}$.

In terms of the product of local 1-forms \(\left(\text{dx}^n\right)\) and
\(\sqrt{|\Lambda|}\), a local volume form \(\left(\omega
^*\right)\) of the pseudo-Riemannian manifold is\cite{szekeres,diff1}:
\begin{equation}
\omega ^*\left(x^1,x^2,\text{...}x^n\right)=\sqrt{-|\Lambda|}\text{dx}^1\wedge
\text{dx}^2\wedge \text{...}\wedge \text{dx}^n
\end{equation}
The effective mass of the local volume can be calculated as,
$\mathit{m}_{vac}\equiv \frac{\Lambda c^2}{8\pi G}V_n$. Because we can assume
homogeneity of the background energy density characterized by the cosmological constant (+$\Lambda
$) this value of the effective mass holds for any number of continuum parcels
(k) with volumes equal to that of the n-ball \(\left(V_n\right)\) for
isotropic spaces. Moreover, for this discussion of the dilation field we
limit ourselves to consider hyperbolic spaces of dimension n = 4, to restrict
the complexity of cosmological implications on the field description. Now that
an expression for the amount of mass per unit parcel is given, we continue our
description of the dilation field as analogous to that of a traveling phonon in
4-dimensional hyperbolic spaces.
Consider an initial position at $\{x^0,x^i\}$, a 4-displacement
$S^{\alpha }$ can be introduced resulting in a dilated configuration of the
continuum with a change in volume for $\Delta V = V^{*}$.
The displacement can be described as a continuous function of parameter ($\eta
$) for $S^{\alpha}(\eta)$, requiring that small gradients of $S$ (strain) follow
the linearized theory of elasticity conveyed in Hooke{'}s law. While large
deformations follow the nonlinear theory of continuum deformations. Sustaining
the generality of the elastic theory, we state that gradients of the
displacement are expressed in terms of the aforementioned spacetime strain
tensor $\varepsilon_{\mu\nu}$ as a function of the parameter
($\eta, \tau$)\cite{rg1}.
Similarly, from equation \ref{dilation}, the field can be written as a smooth
function of ($\eta, \tau $),
\begin{eqnarray}
\mathcal{D}^{\mu\nu}(\eta,\tau)
&=& \frac{\beta\mathcal{R}_{vac}}{2
m_{vac}}\Lambda^{\mu\nu}\left[\Lambda_{\sigma\gamma}\left(E^{\sigma\gamma}(\eta,\tau)-\Lambda^{\sigma\gamma}\right)\right]\nonumber\\
&=&\frac{\beta\mathcal{R}_{vac}}{2
m_{vac}}\Lambda^{\mu\nu}\left[\Lambda_{\sigma\gamma}\varepsilon^{\sigma\gamma}(\eta,\tau)\right]
\end{eqnarray}
where $\mathcal{D}_{\mu\nu}=\mathcal{D}^{\mu\nu}=0$ for $\{\mu \neq \nu\}$ and
the constant $\mathcal{R}_{vac}$ is the intrinsic scalar curvature of the
background spacetime$-\Lambda_{\mu\nu}$. The Lagrangian density, $\mathcal{L}$,
of the dilation field with a background spacetime can be expressed as a
combination of the energy associated with the scalar curvature generated by the
dilation field $\mathcal{L}_R$, the kinetic energy density of the field
$\mathcal{L}_{\mathcal{D}}$, and the mass of the field
$\mathcal{L}_{\mathit{m}}$:
\begin{equation}
\mathcal{L}(\Lambda,\mathcal{D}(\eta,\tau),\nabla
\mathcal{D}(\eta,\tau),\text{...})=\mathcal{L}_R-\mathcal{L}_{\mathcal{D}}-\mathcal{L}_{\mathit{m}}
\end{equation}
Where the respective terms are,
\begin{subequations}
\begin{align}
 \mathcal{L}_R = & \frac{c^4}{16\text{$\pi $G}}\Lambda ^{\mu \nu
 }\tilde{R}_{\mu\nu}\\
 \mathcal{L}_{\mathcal{D}} =& \frac{1}{2}\Lambda ^{\alpha \beta }\nabla
 _{\alpha }\mathcal{D}_{\mu \nu }\nabla _{\beta }\mathcal{D}_{\mu \nu } \\
 \mathcal{L}_{\mathit{m}} =&
 \frac{1}{2}m_{\text{vac}}{}^2\left(\mathcal{D}_{\mu \nu }\right){}^2.
\end{align}
\end{subequations}
Here, $c$ is the proposed speed of propagation for the gravitational field, 
with $\tilde{R}_{\mu \nu }$ and  $\tilde{R}$ as the resultant
Ricci curvature tensor and scalar, respectively, produced by the field.
Before combining terms a parametrization of the covariant derivative in terms of
($\eta$) and proper time ($\tau$) is needed to extract the dynamics produced by
a sought after field equation. This parametrization essentially makes the
covariant derivative a differential operator in terms of the coordinate-free
parameter, proper time, giving us the following langrangian,
\begin{widetext}
\begin{eqnarray}
 \mathcal{L}(\Lambda, \mathcal{D}(\eta,\tau),\nabla
 \mathcal{D}(\eta,\tau))&=&\frac{c^4}{16\text{$\pi$G}}\Lambda^{\mu\nu}\tilde{R}_{\mu\nu}-\frac{1}{2}\Lambda
 ^{\alpha \beta }\nabla _{\alpha }\mathcal{D}_{\mu \nu }\nabla _{\beta }\mathcal{D}_{\mu \nu }-\frac{1}{2}m_{\text{vac}}{}^2\left(\mathcal{D}_{\mu \nu }\right){}^2 \\
 &\equiv&\frac{c^4}{16\text{$\pi $G}}\Lambda ^{\mu \nu
 }\tilde{R}_{\mu\nu}-\frac{1}{2}\left[\left(D_{(\tau)}\mathcal{D}_{\mu\nu}\right){}^2+\left(D_{(\eta)}\mathcal{D}_{\mu \nu }\right){}^2\right]-\frac{1}{2}m_{\text{vac}}{}^2\left(\mathcal{D}_{\mu\nu}\right){}^2\nonumber
\end{eqnarray}
\end{widetext}
where the operators \(D_{(\tau )}=T^{\alpha }\nabla _{\alpha }\), with timelike tangent vector \(T^{\alpha }=\frac{\partial \eta ^{\alpha
}}{\partial \tau }\) and \(D_{(\eta )}=S^{\alpha }\nabla _{\alpha }\), with
spacelike tangent vector \(S^{\alpha }=\frac{\partial }{\partial \eta ^{\alpha
}}\). This separation of timelike and spacelike differential operators is key to
producing evolutions of the tensor field and ultimately generating a
parametrized wave operator\cite{rgdiss}. 

This tensor field is also predicted to interact with the massive
background spacetime and itself, thus we include nonlinear interaction terms in
$\mathcal{L}_{m}$ up to order-$\sigma$ and coupled by a constant
($\mathit{g}_{\sigma}$) that accounts for the strength of the field-field and field-background
coupling.\cite{klein1,klein3,klein4}
\begin{equation}
\mathcal{L}_{\mathit{m}} \to \mathcal{L}_{\mathit{m}}=\frac{1}{2}m_{\text{vac}}{}^2\left(\mathcal{D}_{\mu \nu }\right){}^2+\sum _{\sigma } \frac{1}{\sigma
!}\mathit{g}_{\sigma}\left(\mathcal{D}_{\mu \nu }\right){}^{\sigma }
\end{equation}
The addition of this $\sigma-$interaction gives a formal Lagrangian density as:
\begin{eqnarray}
\mathcal{L}(\Lambda_{\mu\nu},
\mathcal{D}_{\mu\nu},\nabla\mathcal{D}_{\mu\nu})&=&\frac{c^4}{16\text{$\pi$G}}\Lambda^{\mu\nu}\tilde{R}_{\mu\nu}\\\nonumber
&-&\frac{1}{2}[\left(\mathcal{D}_{(\tau)}\mathcal{D}_{\mu\nu}\right){}^2 +\left(\mathcal{D}_{(\eta)}\mathcal{D}_{\mu\nu}\right){}^2]\\\nonumber
&-&\frac{1}{2}m_{\text{vac}}{}^2\left(\mathcal{D}_{\mu\nu
}\right){}^2\\\nonumber 
&+&\sum_{\sigma}
\frac{1}{\sigma!}\mathit{g}\left(\mathcal{D}_{\mu\nu
}\right){}^{\sigma} \nonumber
\end{eqnarray}

\section{The Dilation Field Equation}\label{fieldeqn}
From the lagrangian density outlined in the previous section we can continue
further to find appropriate field equations for the field in question. We can
proceed to write down a functional of the background metric and the field
variable for some boundary $\Sigma$.
\begin{equation}
\mathcal{A}[\mathcal{D}, \Lambda ]=\int_{\Sigma}
\left[\mathcal{L}_R-\mathcal{L}_{\mathcal{D}}-\mathcal{L}_{\mathit{m}}\right]\sqrt{-\Lambda}d^{\mu}\eta
\end{equation}
The extremization of the action ($\mathcal{A}$) is taken with respect to the
field variable ($\bf{\mathcal{D}}$) and also the background metric
$\bf{\Lambda}$. This allows a unique variation of the action,
giving a new interdependency of the lagrangian density on two variables
describing the nature of multiple species of the same gravitational field.
Executing this variation, $\mathcal{A}[\mathcal{D} + \delta \mathcal{D}]$
is
\begin{align}
 \mathcal{A}[\mathcal{D} + \delta \mathcal{D}]&\equiv \mathcal{A}[\mathcal{D}] + \delta\mathcal{A}[\delta
 \mathcal{D}]\\ \nonumber
 &=\int_{\Sigma}
 \left[\mathcal{L}_{\mathcal{D}}+\mathcal{L}_{\mathit{m}}\right]d^{\mu
 }\eta\\\nonumber 
 &+ \int_{\Sigma}
 \biggl[\frac{\partial\left(\mathcal{L}_{\mathcal{D}}+\mathcal{L}_{\mathit{m}}\right)}{\partial\mathcal{D}_{\mu\nu}}\delta\mathcal{D}_{\mu\nu}\\\nonumber
 &+\frac{\partial\left(\mathcal{L}_{\mathcal{D}}+\mathcal{L}_{\mathit{m}}\right)}{\partial
 \left(\partial _{\alpha}\mathcal{D}_{\mu\nu}\right)}\delta
 \left(\partial_{\alpha}\mathcal{D}_{\mu\nu}\right)\biggr]\sqrt{-\Lambda}d^{\mu}\eta\nonumber
 \end{align}
The last term in the above equation is the variation of the action with respect
to the field denoted by $\delta \mathcal{A}[\delta \mathcal{D}]$ and can be
expanded as
\begin{widetext}
\begin{eqnarray}
 \delta \mathcal{A}[\delta \mathcal{D}] &=&\int_{\Sigma }\left[\frac{\partial
 \left(\mathcal{L}_{\mathcal{D}}+\mathcal{L}_{\mathit{m}}\right)}{\partial
 \mathcal{D}_{\mu \nu }}\delta \mathcal{D}_{\mu\nu}
 +\frac{\partial\left(\mathcal{L}_{\mathcal{D}}+\mathcal{L}_{\mathit{m}}\right)}{\partial
 \left(\partial_{\alpha }\mathcal{D}_{\mu\nu}\right)}\delta \left(\partial
 _{\alpha }\mathcal{D}_{\mu\nu}\right)\right]\sqrt{-\Lambda }+d^{\mu
 }\eta\\
 &=&\int _{\Sigma} \left[\frac{\partial
 \left(\mathcal{L}_{\mathcal{D}}+\mathcal{L}_{\mathit{m}}\right)}{\partial \mathcal{D}_{\mu \nu }}-\partial _{\alpha }\frac{\partial \left(\mathcal{L}_{\mathcal{D}}+\mathcal{L}_{\mathit{m}}\right)}{\partial \left(\partial _{\alpha }\mathcal{D}_{\mu \nu
}\right)}\right]\delta \mathcal{D}_{\mu \nu }\sqrt{-\Lambda }d^{\mu }\eta +\delta \mathcal{D}_{\mu \nu }\frac{\partial \left(\mathcal{L}_{\mathcal{D}}+\mathcal{L}_{\mathit{m}}\right)}{\partial
\left(\partial _{\alpha }\mathcal{D}_{\mu \nu }\right)}
\biggm|_{\partial \Sigma }\nonumber
\end{eqnarray}
\end{widetext}
which includes a boundary term at the surface boundary $\partial \Sigma $,
provided that the variation at the fixed boundary yields the following subset for
$\mathcal{L}_{\mathcal{D}}$ and $\mathcal{L}_{\mathit{m}}$ of the
Euler-Lagrange equations of motion
\begin{eqnarray}
 0&=&\int_{\Sigma}
 \biggl[\frac{\partial\left(\mathcal{L}_{\mathcal{D}}+\mathcal{L}_{\mathit{m}}\right)}{\partial\mathcal{D}_{\mu\nu}}\\\nonumber
 &-&\partial_{\alpha}\left(\frac{\partial
 \left(\mathcal{L}_{\mathcal{D}}+\mathcal{L}_{\mathit{m}}\right)}{\partial
 \left(\partial_{\alpha}\mathcal{D}_{\mu \nu}\right)}\right)\biggr]\delta
 \mathcal{D}_{\mu\nu}\sqrt{-\Lambda}d^{\mu}\eta
\end{eqnarray}
Because the Lagrangian density is also dependent on the background metric
$\Lambda ^{\mu \nu }$ through the interaction of the background, we vary the
action once again but with respect to the inverse metric $\Lambda _{\mu
\nu}$. This variation will give us the appropriate equation describing the
curvature of space from the interaction with the dilation field. The variation
in the action is now expressed as $\delta \mathcal{A}$[$\delta \mathcal{D}$, $\delta \Lambda $], where $\delta \mathcal{A}$[$\delta \Lambda $] is the variation of the action with respect to the inverse metric:
\begin{eqnarray}
 \delta \mathcal{A}[\delta \Lambda ] &=&\int_{\Sigma }\frac{\delta
 \mathcal{L}_{\mathit{total}}\sqrt{-\Lambda }}{\delta \Lambda ^{\mu \nu }}d^{\mu
 }\eta
 \\
 &=&\int _{\Sigma } \left[\frac{\delta \left(\mathcal{L}_R +
 \mathcal{L}_{\mathcal{D}} + \mathcal{L}_{\mathit{m}}\right)}{\delta \Lambda
 ^{\mu \nu }}\cdot \left(\frac{\delta \sqrt{-\Lambda }}{\delta \Lambda ^{\mu \nu
 }}\right)\right]d^{\mu }\eta\nonumber
\end{eqnarray}
where the variation in \(\delta \Lambda ^{\mu \nu }\tilde{R}_{\mu \nu
}\) can be expanded in terms of $\frac{\delta \tilde{R}}{\delta \Lambda ^{\mu
\nu }}$ and $\frac{\delta \sqrt{-\Lambda }}{\delta \Lambda ^{\mu \nu }}$, such
that
\begin{equation}
\frac{\delta \left(\Lambda ^{\mu \nu }\tilde{R}_{\mu \nu }\right)}{\delta \Lambda ^{\mu \nu }}=\frac{\delta }{\delta \Lambda ^{\mu \nu }}\left[\left(\delta
\Lambda ^{\mu \nu }\right)\tilde{R}_{\mu \nu }+\Lambda ^{\mu \nu }\left(\delta \tilde{R}_{\mu \nu }\right)\right]
\end{equation}
From the cancellation of \(\delta \Lambda ^{\mu \nu }/\delta \Lambda ^{\mu \nu }=1\), this simplifies to just the trace of the variation of the Ricci
tensor with respect to the metric, $\frac{\Lambda ^{\mu \nu }\delta
\left(\tilde{R}_{\mu \nu }\right)}{\delta \Lambda ^{\mu \nu }}\equiv
\frac{\delta \tilde{R}}{\delta \Lambda ^{\mu \nu }}$. In order to evaluate
$\delta $\(\tilde{R}\), we first evaluate $\delta \tilde{R}_{\mu \nu }$ as a
contraction of the first and second indeces in the respective Riemann curvature
tensor; 
\begin{equation}
\delta \tilde{R}_{\mu \nu }=\delta \tilde{R}^{\rho }{}_{\mu \rho \nu }=\nabla _{\rho }\left(\delta \Gamma ^{\rho }{}_{\nu \mu }\right)-\nabla _{\nu
}\left(\delta \Gamma ^{\rho }{}_{\rho \mu }\right).
\end{equation}
Thus, taking the trace results in obtaining the variation of the scalar
curvature $\delta\tilde{R}$
\begin{eqnarray}
\Lambda ^{\mu \nu}\delta \tilde{R}_{\mu \nu}&=&\delta
\tilde{R}\\\nonumber
&=&\tilde{R}_{\mu \nu }\delta \Lambda ^{\mu \nu}+\nabla _{\gamma}\left(\Lambda
^{\mu \nu}\delta \Gamma ^{\gamma}{}_{\nu \mu}-\Lambda ^{\mu \gamma}\delta \Gamma
^{\rho }{}_{\rho \mu }\right)
\end{eqnarray}
The last term in the above equation becomes a total derivative with respect to
the metric and does not contribute to the variation of the action at the
boundary $\partial \Sigma $. Emulating the expected Einstein field equations,
the resulting variation with respect to the background metric becomes
\begin{equation}
\frac{c^4}{16\pi G}\left(\tilde{R}_{\mu \nu}-\frac{1}{2}\Lambda
_{\mu \nu}\tilde{R}\right)=\frac{c^4}{16\pi G}\tilde{G}_{\mu\nu}
\end{equation}
The result of this elegant variation is what we expect when considering the
initial dilation field scalar curvature, $\tilde{R}$, in the total Lagrangian
density for the field and it's interaction. This subset of the full equations of
motion represents the curvature associated with the massive dilation field. Accounting for a
second source of curvature separate from the original background measure. 

Adding this back into the total Lagrangian density for the dilation field we
have,
\begin{widetext}
\begin{eqnarray}
\delta \mathcal{A}[\delta \mathcal{D}_{\mu\nu},\delta
\Lambda_{\mu\nu}]&=&\int_{\Sigma} \biggl[\biggl[\frac{c^4}{16\text{$\pi $G}}\left(\tilde{R}_{\mu\nu
}-\frac{1}{2}\Lambda_{\mu \nu}\tilde{R}\right)
+\frac{1}{2}\Lambda_{\mu \nu}\left(\mathcal{L}_{\mathcal{D}} + \mathcal{L}_{\mathit{m}}\right)+\frac{\delta
\left(\mathcal{L}_{\mathcal{D}} + \mathcal{L}_{\mathit{m}}\right)}{\delta
\Lambda ^{\mu \nu}}\biggr]\delta \Lambda ^{\mu \nu}\\\nonumber 
&+&\biggl[\left(\frac{\partial\left(\mathcal{L}_{\mathcal{D}}+\mathcal{L}_{\mathit{m}}\right)}{\partial\mathcal{D}_{\mu\nu}}
-\partial_{\alpha}\left(\frac{\partial
\left(\mathcal{L}_{\mathcal{D}}+\mathcal{L}_{\mathit{m}}\right)}{\partial
\left(\partial_{\alpha}\mathcal{D}_{\mu \nu}\right)}\right)\right)\delta
\mathcal{D}_{\mu \nu}\biggr]\biggr]\sqrt{-\Lambda}d^{\mu}\eta \\\nonumber
&=&0
\end{eqnarray}
\end{widetext}
This statement holds true from the variational principle for stationary actions,
stating that $\delta \mathcal{A}$ = 0. Since the variations in \(\delta
\mathcal{D}_{\mu \nu }\) and \(\delta \Lambda _{\mu \nu }\) are arbitrary, the
interior terms for the varied action must vanish appropriately such that the
full Euler-Lagrange equations of motion can be written as:
\begin{widetext}
\begin{equation}
 \left[\frac{c^4}{16\text{$\pi $G}}(\tilde{R}_{\mu \nu }-\frac{1}{2}\Lambda _{\mu \nu }\tilde{R})+\frac{1}{2}\Lambda _{\mu \nu }\left(\mathcal{L}_{\mathcal{D}}
+ \mathcal{L}_{\mathit{m}}\right)+\frac{\delta \left(\mathcal{L}_{\mathcal{D}} +
\mathcal{L}_{\mathit{m}}\right)}{\delta \Lambda ^{\mu \nu }}\right]
+\left[\frac{\partial \left(\mathcal{L}_{\mathcal{D}}+\mathcal{L}_{\mathit{m}}\right)}{\partial \mathcal{D}_{\mu \nu }}-\partial _{\alpha }\left(\frac{\partial \left(\mathcal{L}_{\mathcal{D}}+\mathcal{L}_{\mathit{m}}\right)}{\partial
\left(\partial _{\alpha }\mathcal{D}_{\mu \nu }\right)}\right)\right]=0
\end{equation}
\end{widetext}
Terms in this equation of motion can be compacted by replacing the differential
terms of the Lagrangian density for their respective counterparts in terms of
the field variable. In doing so, this allows us to restate this equation of
motion in a form that is appropriate for a \textit{nonlinear wave equation}.
Identifying the terms as:
\begin{eqnarray}
&&\frac{\partial
\left(\mathcal{L}_{\mathcal{D}}+\mathcal{L}_{\mathit{m}}\right)}{\partial \mathcal{D}_{\mu \nu}}-\partial_{\alpha}\left(\frac{\partial \left(\mathcal{L}_{\mathcal{D}}+\mathcal{L}_{\mathit{m}}\right)}{\partial
\left(\partial_{\alpha}\mathcal{D}_{\mu
\nu}\right)}\right)\\\nonumber 
&=&\left(\Lambda ^{\alpha \beta}\nabla_{\alpha
}\nabla_{\beta}+\mathit{m}_{\text{vac}}{}^2\right)\mathcal{D}_{\mu \nu}+\sum \frac{(\sigma)}{\sigma !}\mathit{g}\left(\mathcal{D}_{\mu \nu}\right){}^{(\sigma -1)}
\end{eqnarray}
and
\begin{eqnarray}
\frac{c^4}{16\pi G}\tilde{R}_{\mu \nu }=
&-&\frac{1}{2}\Lambda_{\mu\nu}\left(\tilde{R}+\mathcal{L}_{\mathcal{D}}
+\mathcal{L}_{\mathit{m}}\right)\\
&+&\frac{\delta \left(\mathcal{L}_{\mathcal{D}} +
\mathcal{L}_{\mathit{m}}\right)}{\delta \Lambda^{\mu \nu}}\nonumber
\end{eqnarray}
Substituting in for the above terms gives a simplified equation of motion for
the dilation field \(\left(\mathcal{D}_{\mu \nu }\right),\)
\begin{eqnarray}
0&=&\left(\Lambda^{\alpha \beta}\nabla_{\alpha}\nabla_{\beta
}+\mathit{m}_{\text{vac}}{}^2\right)\mathcal{D}_{\mu \nu}\\\nonumber
&+&\sum
\frac{(\sigma)}{\sigma !}\mathit{g}\left(\mathcal{D}_{\mu \nu }\right){}^{(\sigma -1)}+\frac{c^4}{16\text{$\pi $G}}\tilde{R}_{\mu \nu}
\end{eqnarray}
The above equation of motion is very much close to the phononic approximation
for a massive longitudinal wave. Replacing the second covariant
derivatives with a modified wave operator $\tilde{\square }$ reminiscent of a
gauge derivative, we arrive at the final form of the Longitudinal Wave equation
for volume deformed spacetimes (where again, $R^{\beta \nu }$ is the Ricci curvature tensor associated with the
sourced background spacetime, and $\tilde{R}_{\mu\nu}$ associated with
the wave):
\begin{equation}
\Lambda^{\alpha \beta }\nabla_{\alpha }\nabla _{\beta } \to \tilde{\square }=\hat{\square }-R^{\beta \nu }
\end{equation}
Thus we have,
\begin{eqnarray}
 0&=&\left(\hat{\square}+\mathit{m}_{\text{vac}}{}^2\right)\mathcal{D}_{\mu
 \nu}-R^{\beta \nu}\mathcal{D}_{\mu \beta}\\\nonumber 
 &+&\sum \frac{(\sigma
 )}{\sigma !}\mathit{g}\left(\mathcal{D}_{\mu \nu}\right){}^{(\sigma -1)}+\frac{c^4}{16\text{$\pi $G}}\tilde{R}_{\mu \nu}
\end{eqnarray}

\section{Discussion}

\subsection{Interpretation of the Dilation Field Equations}

This massive tensor field effectively generates a classical self-interaction
that interacts with the background spacetime, as can be seen by the production
of a second species of scalar curvature along with the quadratic and
higher-order terms in the expanded potential. These accompanying dynamics
provides a means of transporting the information on the scalar curvature
associated with the source distribution of mass-energy. Here a
``self-interaction'' is defined as the coupling of the field with itself,
generating a classical analog of the quantum field theoretic
\textit{self-energy} \cite{scalar1,noether,qg1,phase,sympl}.
A resistive factor, respective of the kinematic properties during propagation,
included in this derivation of the dilation field is given by the effective mass
$\mathit{m}_{\text{vac}}=\frac{\Lambda c^2}{8\pi G}\int_v\sqrt{-\Lambda }d^{\mu
}x_{\mu}$ of the $\Lambda$-vacuum background.  In regards to the wave equation
introduced from the dilation field equations, this generates the sought after
Klein-Gordon like form for the massive longitudinal wave equation. Taking a
closer look at the Lagrangian density of this field reveals that it admits a
behavior similar to what is expected of a massive scalar field.
The nonlinearity, and subsequently the self-interaction, of the field is evident
in the terms that couple the field with itself (field-field interactions) and
couplings with the background $\Lambda^{\mu\nu}$ spacetime (field-vacuum
interactions), represented as:
\begin{equation}
\Omega\left(\mathcal{D}_{\mu\nu}\right)
=\frac{c^4}{16\text{$\pi$G}}\tilde{R}_{\mu\nu}+\sum_{\sigma}^{n}\frac{(\sigma)}{\sigma!}\mathit{g}_{\sigma}\left(\mathcal{D}_{\mu\nu}\right){}^{(\sigma
-1)}
\end{equation}
involving a coupling index ($\sigma$) and strength
constant in ($\mathit{g}$).
Expanding this polynomial for n = 4 gives an expression for the coupling
strength of the scalar-tensor gravitational field, with ($\mathit{g}_{\sigma}$)
the coupling constants. Further constraints on the coupling constants will not be
explored in this text, such that we limit ourselves to just approximating the
form of the interaction polynomial.
\begin{eqnarray}
\sum_{\sigma}^{n=4}\frac{(\sigma)}{\sigma!}\mathit{g}_{\sigma}\left(\mathcal{D}_{\mu\nu}\right){}^{(\sigma
-1)} &=& \mathit{g}_{1} +
\mathit{g}_{2}\left(\mathcal{D}_{\mu\nu}\right){}^{1}\\\nonumber 
&+& \frac{1}{2}\mathit{g}_{3}\left(\mathcal{D}_{\mu\nu}\right){}^{2}
+ \frac{1}{8}\mathit{g}_{4}\left(\mathcal{D}_{\mu\nu}\right){}^{3}
\end{eqnarray}
For an n=4 coupling we can see that only the third and fourth terms contribute
to meaningful couplings of the field for which the proposed self-interactions
are prominent. For the lagrangian density of the field, restated here for convenience,
\begin{eqnarray}
\mathcal{L}(\Lambda_{\mu\nu},
\mathcal{D}_{\mu\nu},\nabla\mathcal{D}_{\mu\nu})
&=&
\frac{1}{2}[\left(\mathcal{D}_{(\tau}\mathcal{D}_{\mu\nu}\right){}^2
-\left(\mathcal{D}_{(\eta)}\mathcal{D}_{\mu\nu}\right){}^2]\nonumber\\ &-&
\frac{1}{2} m^{2}\mathcal{D}_{\mu\nu} + \Omega\left(\mathcal{D}_{\mu\nu}\right)
\end{eqnarray}
the following properties can be stated.
The lagrangian is real if, $m^{2}, \mathcal{D}^{\mu\nu}, g_{\sigma} \in
\mathbb{R}$; giving a stable field theory. The first two terms are quadratic in
$\mathcal{D}^{\mu\nu}$, stating that if $\tilde{R}$, and $g_{\sigma}$ are zero
we have a free-field lagrangian density $(\mathcal{L}_{0})$ described by the
Klein-Gordon equation of motion. Representing a free tensor field theory. This
term represents the free-kinetic component of the lagrangian. Here $m$ is the
classical mass of the field. Foreshadowing a quantization of this lagrangian,
$g_{\sigma}$ is a coupling constant that is a measure of interaction strength
with cross-section proportional to $g^{2}$. Renormalizability can remove all the
infinities of the resulting quantum field theory, provided all coefficients have
units of $[M]^{n}, n \geq 0$. This implies that there are no
$\frac{1}{6!}g_{6}(\mathcal{D}_{\mu\nu})^{6}$ terms appearing in the polynomial
$\Omega(\mathcal{D}_{\mu\nu})$. A cubic term,
$\frac{1}{3!}g_{3}(\mathcal{D}_{\mu\nu})^{3}$, is allowed if and only if the
lagrangian is a Lorentz scalar and $m$ is interpreted as \textit{mass}. This
requires that there be no linear terms of the field variable like that of
$\hat{\mu}(m)\mathcal{D}_{\mu\nu}$. We extend these properties to the kinetic
term,$-\frac{1}{2}\left(\hat{\nabla}_{\alpha}\mathcal{D}_{\mu\nu}\hat{\nabla}^{\alpha}\mathcal{D}_{\mu\nu}\right)$,
for the propagation of the energy-momentum contained in the field. Requiring
that this term is quadratic in the field variable. Under the variation of the
action, as seen in the previous section, this results in the generation of the
wave operator of curved spacetimes,
$\hat{\square}-R^{\beta}_{\nu}=\{\tilde{\square}\}^{\beta}_{\nu}$.

In the field
expression, the Ricci curvature again provides the Einstein field equations governing the local curvature of spacetime, in which the curvature scalar holds values for the background vacuum and the source distribution. When considering the $\Lambda$-vacuum as a zero state configuration, the scalar curvature takes on a unique role for this field.
We consider the following relationships for all \textit{species} of scalar
curvatures
\begin{subequations}
\begin{eqnarray}
R_{vac}[\Lambda^{\alpha\sigma}] &=&
\Lambda^{\alpha\sigma}R_{\alpha\sigma}\propto 4\Lambda + R_{source}\\
\tilde{R}[\varepsilon^{\alpha\sigma}] &=&
\Lambda^{\alpha\sigma}\tilde{R}_{\alpha\sigma} \propto \Gamma (\varepsilon)
\end{eqnarray}
\end{subequations}
We can see that once the field equations are constructed, there exists two
distinctly separate scalar curvatures that are unique to the background
spacetime $\left(R_{vac}[\Lambda^{\alpha\sigma}]\right)$ and the dilation field
($\tilde{R}[\varepsilon^{\alpha\sigma}]$), respectively. With the
explanation of a unique variation of the action given in the previous
section, one can use this explicit determination of scalar curvature
\textit{species} to make a statement on a proposed gravitational charge. One
can see that in the expression for explicitly defining the field variable,
$\mathcal{D}^{\mu\nu}$, the scalar curvature behaves like that of the ``charge''
of the field. With this, one can make a substantial comparison to the behavior
of a source charge generating a field. Considering the scalar curvature
generated as a source term for the field we can state a massive longitudinal
wave equation that gives variations in the volumes of constituent finite spacetime parcels with zero spacing between them. Analogous to the model of a lattice phonon as the vibratory motion of a
particle continuum, the dilation field has been derived much like the
phononic model with appropriate modifications for a
\textit{relativistic classical field} in four spacetime dimensions.
With this, spacetime fluidity is approximated by a collection of local unit
volumes with mass ($\mathit{m}_{\text{vac}}$) and measure of persistence
($\beta$), as previously stated. This foundation provides the massive
longitudinal wave equation in terms of the field tensor giving the
configuration at unit time:
\begin{eqnarray}
 0&=&\left(\hat{\square}+\mathit{m}_{\text{vac}}{}^2\right)\mathcal{D}_{\mu
 \nu}-R^{\beta \nu}\mathcal{D}_{\mu \beta}\\\nonumber 
 &+&\sum \frac{(\sigma
 )}{\sigma !}\mathit{g}\left(\mathcal{D}_{\mu \nu}\right){}^{(\sigma -1)}+\frac{c^4}{16\text{$\pi $G}}\tilde{R}_{\mu \nu}
\end{eqnarray} 

\subsection{Dispersion Relation Approximation and Effective Vacuum Mass}
With a formal expression of the complete wave equations we are now able to
discuss a dispersion relation for the propagating longitudinal wave.
Again, in comparison to that of the electromagnetic field radiation, the EM
dispersion relation states that the inner product of the 4-wave vector
($K_{\mu}K^{\mu}=0$) returns a null value. Resulting in the confirmation that
electromagnetic waves, as transverse waves, propagate at the speed of light in
vacuum\cite{maxwell1}. This statement can be generally applied to all massless
transverse waves in vacuum giving this lightlike dispersion. In contrast to
this, longitudinal waves, requiring they have some representation of a
``medium'' to propagate through, have a different dispersion relation. Under the
analogy to seismic pressure waves, due to the density of the medium in which p-waves propagate through, these longitudinal
waves travel at speeds greater than their transverse counterparts
(shear-waves)\cite{disp1}. With this dissimilarity in mind, it has now been
confirmed through observation that transverse gravitational waves propagate at the speed of light
giving them a null dispersion relation. As for longitudinal gravitational waves
propagating through the ``medium'' that is the massive isotropic vacuum, their
dispersion relation should reveal that these waves propagate at superluminal
speeds. Generating a contrasting result for their dispersion relation in
comparison to their transverse counterparts. For the 4-wave vector having a
spacetime signature (+,-,-,-), it can be defined as
\begin{equation}
K_{\mu} = \left(\frac{\omega}{c_{g}}, -K_{i}\right)
\end{equation}
where ($c_{g}$) is the group velocity of the longitudinal wave. We can make a
substitution for the group velocity and restate the 4-wave vector for
longitudinal waves as
\begin{equation}
K_{\mu} = \left(\sqrt{\frac{\rho_{vac}}{\beta}}\omega, -K_{i}\right)
\end{equation}
Taking the inner product of the wave vector gives
\begin{eqnarray}
K_{\mu}K^{\mu}&=&\left(\sqrt{\frac{\rho_{vac}}{\beta}}\omega\right)^2 -
K_{i}K^{i}\\
&=& \frac{\rho_{vac}}{\beta}\omega^2 - K_{i}K^{i}\nonumber\\
&=& \frac{\rho_{vac}}{\beta}\omega^2
-\left(\frac{\omega}{v_{p}}\right)^2\hat{n}_{i}\hat{n}^{i}\nonumber
\end{eqnarray}
The inner product of the spatial components are represented as the inner product
of the spatial unit vector ($\hat{n}_{i}$) multiplied by the square of the
angular frequency over the phase velocity. In the massive vacuum the assumption
that the phase velocity equal in magnitude the group velocity cannot be stated
in this case for longitudinal gravitational waves. We can employ the dispersion
relation for the longitudinal wave coming from a rotating compact dense object
with the magnitude of the angular velocity equating to the angular frequency of
the radiated wave. With this, the modulus for bulk gravity is given by the
aforementioned derivation using the Friedmann equations\cite{cosmo1}, where
($\beta$) is:
\begin{equation}
\beta =-\frac{\sqrt{-\det[\Lambda_{\mu\nu}]}}{\alpha ^38\text{$\pi
$G}}\left(3H^2c^2-\text{$\Lambda $c}^4\right)
\end{equation}
This equation for the dilation field represents a way of dynamically describing
the gravitational strain field generated by a nonzero mass-energy distribution
using the trace Einstein field equations. Relativistically, a linear
longitudinal wave is comprised of oscillations parallel to the direction of propagation
resulting in the relaxation (expansion) and rarefaction (compression) of a local
volume. For an N-dimensional spacetime manifold, the dilation field tensor,
$\mathcal{D}_{\mu \nu }$. While the bulk modulus is characterized by a
description of the deformed volume\cite{rg1}.

The effective mass of the local volume can be calculated as,
$\mathit{m}_{\text{vac}}\equiv \frac{\Lambda c^2}{8\pi G}V_n|_{n=4}$. Because we can assume homogeneity of the background energy density
characterized by the cosmological constant (+$\Lambda $) this value of the effective mass holds for any number of continuum parcels (k) with volumes equal to that of the unit n-ball, \(\left(V_n\right)\), for isotropic spaces. Furthermore, for this discussion of the dilation field we limit ourselves to isotropic spaces of dimension n = 4, to restrict the complexity
of the field description. Now that an expression for the amount of mass per unit parcel is given, we continue our description of the dilation field
as analogous to that of a traveling phonon.
For an initial position \(\left(\mathcal{X}^{\alpha }\right)\) with coordinates \(\left\{x^0,x^i\right\}\), a displacement \(\left(S^{\alpha }\right)\)
can be applied resulting in a dilated configuration of the continuum with a
change in volume for \(\left(\Delta V = V^*\right)\). An estimation of the
classical value of the \textit{effective} mass of the gravitational vacuum.
Where ($ G =6.67408\times10{}^{-11} N \cdot kg^{2}/m^{2}$) is the Newtonian
gravitational constant, ($c = 299,792,458\hspace{5pt} m/s^{2}$) the speed of
light in vacuum, ($\Lambda = 2.036\times10{}^{-52}\hspace{5pt} m^{-2}$) the
cosmological constant (not to be confused with the strong-form background metric
tensor, $\Lambda_{\mu\nu}$), and the critical density of the universe
($\rho_{\text{crit}}=9.3\times10^{-27}\hspace{5pt} kg/m^{3}$). The density of
the vacuum can be approximated by using the percentage of the critical density that is attributed to the $\Lambda$-CDM vacuum density
($\rho_\text{vac}$); $\rho_{\text{vac}} = 0.728\times\rho_{\text{crit}} =
6.7704\times10^{-27}$.
We can convert this energy density to an effective \textit{mass} density with a
proper handling of the speed of light (c) by way of the equivalence principle.
Giving an approximate effective mass density of the vacuum to be:
$\rho_{m-vac} = 6.0849\times10^{-10}$. Numerically integrating this
expression bounded by the unit volume for a cartesian coordinate system gives an effective amount of mass contained in a
unit cube in three spatial dimensions $m_{vac} = 6.0849\times10^{-10} kg$.
A quick approximation shows that given the inner product above the dispersion
relation for the longitudinal wave gives a non-trivial result. This tells us
just a small sample of the full derived dispersion relation for the massive
vacuum. Due to the nonlinearity of the longitudinal gravitational wave, after
further investigation, we can expect a dispersion relation that would compliment
this nonlinearity. Further discussion of this relationship will be given in
subsequent work that further investigates this connection to the vacuum density.

\bibliographystyle{apsrev4-1}
\bibliography{REF_full}


\end{document}